\documentclass[12pt]{article}
\usepackage{amsmath}
\usepackage{amssymb}
\usepackage{amsfonts}
\usepackage{latexsym}
\usepackage{color}
\usepackage{graphicx}

\catcode `\@=11 \@addtoreset{equation}{section}

\catcode `\@=12

\newcommand{\be}{\begin{equation}}
\newcommand{\en}{\end{equation}}
\newcommand{\bea}{\begin{eqnarray}}
\newcommand{\ena}{\end{eqnarray}}
\newcommand{\beano}{\begin{eqnarray*}}
\newcommand{\enano}{\end{eqnarray*}}
\newcommand{\bee}{\begin{enumerate}}
\newcommand{\ene}{\end{enumerate}}

\newcommand{\Hil}{{\cal H}}

\newcommand{\Id}{1\!\!1}
\newcommand{\F}{{\cal F}}

\newcommand{\E}{{\cal E}}

\newcommand{\1}{1 \!\!\! 1}
\newtheorem{thm}{Theorem}

\newtheorem{prop}[thm]{Proposition}
\newtheorem{defn}[thm]{Definition}

\begin{document}

\thispagestyle{empty}

\vspace*{1cm}

\begin{center}
{\Large \bf Vector coherent states and intertwining operators}   \vspace{2cm}\\

{\large F. Bagarello}\\
  Dipartimento di Metodi e Modelli Matematici,
Facolt\`a di Ingegneria, Universit\`a di Palermo, I-90128  Palermo, Italy\\
e-mail: bagarell@unipa.it
\end{center}


\vspace*{2cm}

\begin{abstract}
\noindent In this paper we discuss a general strategy to construct vector coherent states of the Gazeau-Klauder type and we use them to built up examples of isospectral hamiltonians. For that we use a general strategy recently proposed by the author and which extends well known facts on intertwining operators. We also discuss the possibility of constructing non-isospectral hamiltonians with related eigenstates.

\end{abstract}

\vspace{2cm}

\vfill

\newpage

\section{Introduction}

In a recent paper, \cite{bag}, we have proposed a new procedure which gives rise, given few ingredients, to a  hamiltonian $h_2$ which has the same spectrum of a given hamiltonian $h_1$ and whose respective eigenstates are related  by a given intertwining operator. These results extend what was discussed in the previous literature on this subject, \cite{intop}, and have the advantage of being a constructive procedure: while in \cite{intop} the existence of $h_1, h_2$ and of an operator $x$ satisfying the rule $h_1x=xh_2$ is assumed, in \cite{bag} we explicitly construct $h_2$ from $h_1$ and $x$ in such a way that $h_2$ satisfies a {\em weak form} of the intertwining condition $h_1x=xh_2$. Moreover, $h_2$ has the same spectrum of $h_1$ and the eigenvectors are related in a standard way, see \cite{bag} and Section III. It is well known that this procedure is strongly related to, and actually extends, the supersymmetric quantum mechanics widely discussed in the past years, see \cite{CKS} and \cite{jun} for interesting  reviews.

In \cite{bag} we have considered the  relation between this interwining operator technique and vector Gazeau-Klauder like coherent states (VGKCS), going in {\em one direction}. Here we continue this analysis showing that the opposite can be done. Namely, we will first introduce two different classes of VGKCS. Their properties are discussed in Section II. In Section III we will show that, starting from these states, several isospectral hamiltonians can be defined. Many examples are discussed, and some of them remind us of supersimmetric quantum mechanics. In Section IV we discuss the possibility of using the same strategy proposed in \cite{bag} to construct non-isospectral hamiltonians whose eigenvectors are related as in \cite{bag}-\cite{jun}. Section V contains our conclusions and future plans.

\section{Vector coherent states}

In this section we extend the framework discussed in \cite{bag} and used there to construct a certain type of coherent states (CS). As we have discussed in \cite{bag}, there is not an unique way to do this. On the contrary,  in the literature  several possibilities are discussed, see \cite{cjt,gk,alibag,fhro,aleixo,AEG,sanali} and references therein. These differences arise mainly because of the non-uniqueness of the definition of what a CS should be. To be more explicit, while some author defines them as eigenvectors of some sort of annihilation operators, \cite{cjt}, someone else appears more interested in getting a resolution of the identity, \cite{gk}. Also the {\em domain} of the CS plays an important role: while for standard CS the domain is a (subset) of $\Bbb{C}$, for vector CS (VCS) the domain is a suitable set of matrices, \cite{AEG,sanali}. It should also be mentioned that VCS were introduced in a different context in \cite{rowe} in connection with group representation theory.

Here, as in \cite{bag}, we adopt a {\em mixed} point of view, showing how to  generalize the Gazeau-Klauder (GK) scheme, \cite{gk}, to the Ali and coworker settings, \cite{AEG}, getting  VCS which still share with the GK ones most of their features. To keep the paper self-contained we first briefly recall how these states are defined and which are their main properties. These CS, labeled by two
parameters $J>0$ and $\gamma\in\Bbb{R}$, can be written in terms of the o.n.
basis of a self-adjoint operator $H=H^\dagger$, $|n>$, as \be
|J,\gamma>=N(J)^{-1}\,\sum_{n=0}^\infty\,\frac{\,J^{n/2}\,e^{-i\epsilon_n\,\gamma}}{\sqrt{\rho_n}}\,|n>,\label{21}\en
where $0=\epsilon_0<\epsilon_1<\epsilon_2<\ldots$, $\rho_n=\epsilon_n!:=\epsilon_1\cdots\epsilon_n$, $\epsilon_0!=1$,
$H|n>=\omega\,\epsilon_n\,|n>$ and
$N(J)^2=\sum_{n=0}^\infty\,\frac{\,J^{n}\,}{\rho_n}$, which converges for $0\leq J<R$, $R=\lim_n \epsilon_n$ (which could be infinite).  These
states are {\em temporarily
stable}: $e^{-iHt}\,|J,\gamma>=|J,\gamma+\omega t>$, $\forall
t\in\mathbb{R}$, and continuous: if $(J,\gamma)\rightarrow(J_0,\gamma_0)$
then $\||J,\gamma>-|J_0,\gamma_0>\|\rightarrow 0$. Moreover they satisfy the {\em action identity}: $<J,\gamma|H|J,\gamma>=J\,\omega$ and a resolution of the identity in the following sense:
if there exists a non negative function, $\rho(u)$, such
that $\int_0^R\,\rho(u)\,u^n\,du=\rho_n$ for all $n\geq 0$ then, introducing a
measure $d\nu(J,\gamma)=N(J)^2\,\rho(J)\,dJ\,d\nu(\gamma)$, with
$\int_{\mathbb{R}}\ldots
\,d\nu(\gamma)=\lim_{\Gamma\rightarrow\infty}\,\frac{1}{2\Gamma}\,
\int_{-\Gamma}^\Gamma\ldots\,d\gamma$, the following holds \be
\int_{C_R}\,d\nu(J,\gamma)\,|J,\gamma><J,\gamma|=\int_0^R\,N(J)^2\,\rho(J)\,dJ\,
\int_{\mathbb{R}}\,d\nu(\gamma)\,|J,\gamma><J,\gamma|=\Id,
\label{22}\en
where $\Id$ is the identity operator. The states $|J,\gamma>$ are
eigenstates of the following $\gamma-$ depending annihilation-like
operator $a_\gamma$ defined on $|n>$ as follows: \be
a_\gamma\,|n>=\left\{
    \begin{array}{ll}
        0,\hspace{4.6cm}\mbox{ if } n=0,  \\
        \sqrt{\epsilon_n}\,e^{i(\epsilon_n-\epsilon_{n-1})\,\gamma}|n-1>, \hspace{0.6cm} \mbox{ if } n>0,\\
       \end{array}
        \right.
\label{23}\en whose adjoint acts as
$a_\gamma^\dagger\,|n>=\sqrt{\epsilon_{n+1}}\,e^{-i(\epsilon_{n+1}-\epsilon_{n})\,\gamma}|n+1>$. We easily deduce that $ a_\gamma |J,\gamma>=\sqrt{J}\,
|J,\gamma>$, even if $|J,\gamma>$ is not an eigenstate of $a_{\gamma'}$ if
$\gamma\neq\gamma'$.

\vspace{2mm}

Let us now consider two self-adjoint hamiltonians $h_1$ and $h_2$, with eigenvalues $\epsilon_n^{(j)}$ and eigenvectors  $\varphi_n^{(j)}$:
\be
h_j\varphi_n^{(j)}=\epsilon_n^{(j)}\varphi_n^{(j)}, \qquad j=1,2, \quad n=0,1,2,\ldots
\label{24}\en
We assume that $0=\epsilon_0^{(j)}<\epsilon_1^{(j)}<\epsilon_2^{(j)}<\cdots$, $j=1,2$. We define \be
  \hat\Psi_n^{(b)}=\left(
                                     \begin{array}{c}
                                       \hat\varphi_n^{(1)} \\
                                       0 \\
                                     \end{array}
                                   \right), \quad \hat\Psi_n^{(f)}=\left(
                                     \begin{array}{c}
                                       0 \\
                                       \hat\varphi_n^{(2)} \\
                                     \end{array}
                                   \right),
                                   \label{25}\en
  where we use, following the same notation as in \cite{alibag,bag} and with a little abuse of language, "b" for bosons and "f" for fermions.
  The set $\F=\{\hat\Psi_n^{(f)},\, \hat\Psi_n^{(b)}, \,n\geq 0\}$ forms an orthonormal basis for
the Hilbert space $\Hil_{susy}:=\Bbb{C}^2\otimes\Hil$, whose scalar product is defined as follows: given $\Gamma=\left(
                                     \begin{array}{c}
                                       \gamma^{(b)} \\
                                       \gamma^{(f)} \\
                                     \end{array}
                                   \right)$ and $\tilde\Gamma=\left(
                                     \begin{array}{c}
                                       \tilde\gamma^{(b)} \\
                                       \tilde\gamma^{(f)} \\
                                     \end{array}
                                   \right)$,  we put $<\Gamma,\tilde\Gamma>_{susy}=<\gamma^{(b)},\tilde\gamma^{(b)}>+<\gamma^{(f)},
                                   \tilde\gamma^{(f)}>$, where $<,>$ is the scalar product in $\Hil$.

Let now $J_1$ and $J_2$ be two positive quantities, $\underline J=(J_1,J_2)$ and  $\gamma$  a real variable. Let further $\delta$ be a strictly positive parameter.  Extending what we have done in \cite{bag} we put
\be
\Psi_\delta(\underline J,\gamma):=\frac{1}{\sqrt{N(\underline J)}}\sum_{n=0}^\infty\,\,\left(\frac{1}{\sqrt{\epsilon_n^{(1)}!}}J_1^{n/2}\,e^{-i(\epsilon_n^{(1)}+\delta)\gamma}\,\hat\Psi_n^{(b)}+
\frac{1}{\sqrt{\epsilon_n^{(2)}!}}J_2^{n/2}\,e^{i(\epsilon_n^{(2)}+\delta)\gamma}\,
\hat\Psi_n^{(f)}\right).
\label{26}\en
With respect to what was done in \cite{bag} here we are {\em doubling} the set of eigenvalues, in the sense that we are not assuming, as it is usually done in the literature so far, that we are dealing with two different operators with the same spectra. Hence $h_1$ and $h_2$ need not to be related, in particular, as in SUSY quantum mechanics or in the ordinary theory of intertwining operators.

The normalization constant $N(\underline J)$ can be easily found requiring as usual that $<\Psi_\delta(\underline J,\gamma),\Psi_\delta(\underline J,\gamma)>_{susy}=1$ for all $\underline J, \gamma$ and $\delta$. Let us define $M_j(J):=\sum_{k=0}^\infty\,\frac{J^k}{\epsilon_k^{(j)}!}$, which converges for $0\leq J<R_j$, $R_j=\lim_n \epsilon_n^{(j)}$, which is assumed to exist (but it could be infinite), $j=1,2$. Then we deduce that
\be
N(\underline J)=M_1(J_1)+M_2(J_2),\label{27}\en
It may be worth remarking that, with respect to \cite{bag}, we have introduced a minor difference in the normalization which, however, does not affect the main results and conclusions. If we now introduce the operator $$
H= \left(
        \begin{array}{cc}
          h_1 & 0 \\
          0 & h_2 \\
        \end{array}
      \right)
$$
acting on $\Hil$,  we find that the following {\em action identity} holds:
\be
<\Psi_\delta(\underline J,\gamma),H\Psi_\delta(\underline J,\gamma)>_{susy}=\frac{J_1M_1(J_1)+J_2M_2(J_2)}{M_1(J_1)+M_2(J_2)}.
\label{28}\en
As for the temporal stability, let us define the matrix
$$
V_\delta(t) = \left(
        \begin{array}{cc}
          e^{-i(h_1+\delta)t} & 0 \\
          0 & e^{i(h_2+\delta)t} \\
        \end{array}
      \right),
$$
then
\be
V_\delta(t)\Psi_\delta(\underline J,\gamma)=\Psi_\delta(\underline J,\gamma+t),
\label{29}\en
for each fixed $\delta$. This means that, independently of $\delta$, $V_\delta(t)$ leaves invariant the set of the vectors in (\ref{26}). As we have already discussed in \cite{bag}, the operator $V_\delta(t)$ does not coincide with $e^{-iHt}$, and for this reason, calling (\ref{29}) {\em temporal stability } is a little abuse of language. We will show in the next subsection that we can avoid such an abuse introducing an extra requirement on the spectra of $h_1$ and $h_2$, $\sigma(h_j)$, $j=1,2$. We leave to the reader to check that the resolution of the identity  can be recovered if we define a measure $d\nu(\underline J,\gamma)$ as follows: $d\nu(\underline J,\gamma)=N(\underline J)\rho_1(J_1)dJ_1\,\rho_2(J_2)dJ_2\,d\nu(\gamma)$, where $\rho_1(J)$ and $\rho_2(J)$ are two non-negative functions satisfying the equality $\int_0^{R_j}\,\rho_j(J)\,J^k\,dJ=\epsilon_k^{(j)}!$, $\forall k\geq 0$. The measure $d\nu(\gamma)$ is defined as usual, see \cite{gk}. With these definitions it is possible to deduce that, for all fixed $\delta>0$,
\be
\int_\E \,d\nu(\underline J,\gamma)\,|\Psi_\delta(\underline J,\gamma)><\Psi_\delta(\underline J,\gamma)|=\Id_{susy},
\label{210}\en
where $\E=\{(\underline J,\gamma): 0\leq J_1<R_1,\,0\leq J_2<R_2,\,\gamma\in\Bbb{R}\}$. As in \cite{bag}, the role of the positive $\delta$ is crucial. Moreover, the integral above is not uniformly continuous in $\delta$, since, if $\delta=0$, it is easy to check that $\int_\E \,d\nu(\underline J,\gamma)\,|\Psi_\delta(\underline J,\gamma)><\Psi_\delta(\underline J,\gamma)|\neq\Id_{susy}$.

Also in this context it is possible to introduce a $\gamma$-depending annihilation like operator. Let
\be
A_\gamma\hat\Psi_n^{(b)}=\left\{
\begin{array}{ll}
0 \hspace{40mm}\mbox{ if }  n=0\\
\sqrt{\epsilon_n^{(1)}}\,e^{i(\epsilon_n^{(1)}-\epsilon_{n-1}^{(1)})\gamma}\hat\Psi_{n-1}^{(b)}\hspace{7mm}\mbox{ if } n\geq 1\\
\end{array}
\right.\label{211}\en
and
\be
A_\gamma\hat\Psi_n^{(f)}=\left\{
\begin{array}{ll}
0 \hspace{40mm}\mbox{ if }  n=0\\
\sqrt{\epsilon_n^{(2)}}\,e^{-i(\epsilon_n^{(2)}-\epsilon_{n-1}^{(2)})\gamma}\hat\Psi_{n-1}^{(f)}\hspace{5mm}\mbox{ if } n\geq 1\\
\end{array}
\right.\label{212}\en
Then the adjoint $A^\dagger_\gamma$ satisfies the following:
\be
\left\{
\begin{array}{ll}
A_\gamma^\dagger\hat\Psi_n^{(b)}=
\sqrt{\epsilon_{n+1}^{(1)}}\,e^{-i(\epsilon_{n+1}^{(1)}-\epsilon_{n}^{(1)})\gamma}\hat\Psi_{n+1}^{(b)}\\
A_\gamma^\dagger\hat\Psi_n^{(f)}=
\sqrt{\epsilon_{n+1}}\,e^{i(\epsilon_{n+1}^{(2)}-\epsilon_{n}^{(2)})\gamma}\hat\Psi_{n+1}^{(f)}\\
\end{array}
\right.\label{213}\en
Then the states $\Psi_\delta(\underline J,\gamma)$ are {\em eigenstates} of the operator $A_\gamma$ in the following sense:
\be
A_\gamma\Psi_\delta(\underline J,\gamma)=J^{1/2}\Psi_\delta(\underline J,\gamma)\label{214o}\en
for all fixed $\delta$, where $J^{1/2}$ is the matrix
$J^{1/2} = \left(
        \begin{array}{cc}
          \sqrt{J_1} & 0 \\
          0 & \sqrt{J_2} \\
        \end{array}
      \right)$.
Hence, the  vectors $\Psi_\delta(\underline J,\gamma)$ can be safely called {\em coherent states}.

\subsection{More assumptions...more results}

The presence of the parameter $\delta$ in the definition (\ref{26}) of the VCS, and of the related operators, may look
a bit unnatural, since is an ad hoc quantity which is used mainly to recover a resolution of the identity. Here we will show that, under a reasonable assumption of the eigenvalues of the two hamiltonians, no $\delta$ is needed.

Once again we consider two hamiltonians $h_1$ and $h_2$ with eigenvalues $\epsilon_n^{(j)}$ and eigenvectors $\varphi_n^{(j)}$:
$
h_j\varphi_n^{(j)}=\epsilon_n^{(j)}\varphi_n^{(j)},$ $j=1,2$, $n=0,1,2,\ldots$.
Contrarily to what we have done before, we assume now that $0<\epsilon_0^{(j)}<\epsilon_1^{(j)}<\epsilon_2^{(j)}<\cdots$, $j=1,2$ and we introduce the following

\begin{defn} $h_1$ and $h_2$ have essentially disjoint spectra (EDS) if $\epsilon_n^{(1)}\neq \epsilon_m^{(2)}$, for all $n$ and $m$ in $\Bbb{N}_0$.
\end{defn}

Of course, this requirement is not compatible with what has been required previously, namely that $0=\epsilon_0^{(1)}=\epsilon_0^{(2)}$, so that this requirement has been removed here.

If $h_1$ and $h_2$ have EDS we can define, using the same notation as before
\be
\Psi(\underline J,\gamma):=\frac{1}{\sqrt{\tilde N(\underline J)}}\sum_{n=0}^\infty\,\,\left(\frac{1}{\sqrt{\tilde\epsilon_n^{(1)}!}}J_1^{n/2}\,e^{-i\epsilon_n^{(1)}\gamma}\,\hat\Psi_n^{(b)}+
\frac{1}{\sqrt{\tilde \epsilon_n^{(2)}!}}J_2^{n/2}\,e^{-i\epsilon_n^{(2)}\gamma}\,
\hat\Psi_n^{(f)}\right),
\label{214}\en
where we have introduced $\tilde\epsilon_n^{(j)}=\epsilon_n^{(j)}-\epsilon_0^{(j)}$, $j=1,2$. Hence $\tilde\epsilon_0^{(j)}!=1$ and $\tilde\epsilon_n^{(j)}!=(\epsilon_n^{(j)}-\epsilon_0^{(j)})(\epsilon_{n-1}^{(j)}-\epsilon_0^{(j)})\cdots(\epsilon_1^{(j)}-\epsilon_0^{(j)})$.
Notice that no $\delta$ is introduced and, furthermore, the two exponentials share the same minus sign. This has interesting consequences on the temporal stability, as we will see in a moment.

The normalization $\tilde N(\underline J)$ can be found as before: let us define $\tilde M_j(J):=\sum_{k=o}^\infty\,\frac{J^k}{\tilde\epsilon_k^{(j)}!}$, which converges for $0\leq J<\tilde R_j$, $\tilde R_j=\lim_n \tilde\epsilon_n^{(j)}$, which is assumed to exist (but it could be infinite), $j=1,2$. Then we deduce that
\be
\tilde N(\underline J)=\tilde M_1(J_1)+\tilde M_2(J_2),\label{215}\en
Rather than computing $<\Psi(\underline J,\gamma),H\Psi(\underline J,\gamma)>_{susy}$, it is more convenient to introduce  a {\em shifted} hamiltonian $H_\tau=\left(
        \begin{array}{cc}
          h_1 & 0 \\
          0 & h_2 \\
        \end{array}
      \right)-\left(
        \begin{array}{cc}
          \epsilon_0^{(1)} & 0 \\
          0 & \epsilon_0^{(2)} \\
        \end{array}
      \right)=:H-\epsilon_0$.
Then we have $H_\tau\hat\Psi_n^{(b)}=\tilde\epsilon_n^{(1)}\hat\Psi_n^{(b)}$ and $H_\tau\hat\Psi_n^{(f)}=\tilde\epsilon_n^{(2)}\hat\Psi_n^{(f)}$. Hence
\be
<\Psi(\underline J,\gamma),H_\tau\Psi(\underline J,\gamma)>_{susy}=\frac{J_1\tilde M_1(J_1)+J_2\tilde M_2(J_2)}{\tilde M_1(J_1)+\tilde M_2(J_2)},
\label{216}\en
which is our version of the action identity.

One of the free bonus that we get using the VCS in (\ref{214}) is that the temporal stability, which is just a formal formula for the states in (\ref{26}), has now a clear physical interpretation: because of the definition of $H$, the time operator $e^{-iHt}$ in $\Hil_{susy}$ is the following two by two matrix:
$$
e^{-iHt} = \left(
        \begin{array}{cc}
          e^{-ih_1t} & 0 \\
          0 & e^{-ih_2t} \\
        \end{array}
      \right),
$$
and it is an easy exercise to check that
\be
e^{-iHt}\Psi(\underline J,\gamma)=\Psi(\underline J,\gamma+t),
\label{217}\en
as expected and as originally deduced in \cite{gk}. The resolution of the identity holds as in the previous case, but a crucial role is played here from the assumption on the spectra of $h_1$ and $h_2$. More explicitly, we put $d\nu(\underline J,\gamma)=\tilde N(\underline J)\rho_1(J_1)dJ_1\,\rho_2(J_2)dJ_2\,d\nu(\gamma)$, where $\rho_1(J)$ and $\rho_2(J)$ are two non-negative functions satisfying the equality $\int_0^{\tilde R_j}\,\rho_j(J)\,J^k\,dJ=\tilde\epsilon_k^{(j)}!$, $\forall k\geq 0$, and  $d\nu(\gamma)$ is defined as before. Furthermore we put $\tilde\E=\{(\underline J,\gamma): 0\leq J_1<\tilde R_1,\,0\leq J_2<\tilde R_2,\,\gamma\in\Bbb{R}\}$. Then we get
\be
\int_{\tilde\E} \,d\nu(\underline J,\gamma)\,|\Psi(\underline J,\gamma)><\Psi(\underline J,\gamma)|=\Id_{susy}.
\label{218}\en
It is clear that we have now no problem of continuity, here, since no parameter $\delta$ appears here. This result directly follows from the assumption that $h_1$ and $h_2$ have EDS, and it would not be true otherwise.

The definitions in (\ref{211})-(\ref{213}) must be slightly modified in our new context:
we put
\be
\tilde A_\gamma\hat\Psi_n^{(b)}=\left\{
\begin{array}{ll}
0 \hspace{44mm}\mbox{ if }  n=0\\
\sqrt{\tilde\epsilon_n^{(1)}}\,e^{i(\epsilon_n^{(1)}-\epsilon_{n-1}^{(1)})\gamma}\hat\Psi_{n-1}^{(b)}\hspace{7mm}\mbox{ if } n\geq 1\\
\end{array}
\right.\label{219}\en
and
\be
\tilde A_\gamma\hat\Psi_n^{(f)}=\left\{
\begin{array}{ll}
0 \hspace{44mm}\mbox{ if }  n=0\\
\sqrt{\tilde\epsilon_n^{(2)}}\,e^{i(\epsilon_n^{(2)}-\epsilon_{n-1}^{(2)})\gamma}\hat\Psi_{n-1}^{(f)}\hspace{5mm}\mbox{ if } n\geq 1\\
\end{array}
\right.\label{220}\en
Then the adjoint $A^\dagger_\gamma$ satisfies the following:
\be
\left\{
\begin{array}{ll}
\tilde A_\gamma^\dagger\hat\Psi_n^{(b)}=
\sqrt{\tilde\epsilon_{n+1}^{(1)}}\,e^{-i(\epsilon_{n+1}^{(1)}-\epsilon_{n}^{(1)})\gamma}\hat\Psi_{n+1}^{(b)}\\
\tilde A_\gamma^\dagger\hat\Psi_n^{(f)}=
\sqrt{\tilde \epsilon_{n+1}^{(2)}}\,e^{-i(\epsilon_{n+1}^{(2)}-\epsilon_{n}^{(2)})\gamma}\hat\Psi_{n+1}^{(f)}\\
\end{array}
\right.\label{221}\en
Again we get
\be
\tilde A_\gamma\Psi(\underline J,\gamma)=J^{1/2}\Psi(\underline J,\gamma).\label{222}\en
In other words, we can get rid of $\delta$ as far as $h_1$ and $h_2$ have EDS, recovering exactly the same properties as before.
Furthermore, it is just an exercise to extend these results to a family of $N$ hamiltonians $h_j$, $j=1,2,\ldots,N$, with EDS (i.e. with all their eigenvalues mutually different). In this case, clearly, we can construct VGKCS in the Hilbert space $\hat\Hil_{susy}:=\Bbb{C}^N\otimes\Hil$. The details of this construction are left to the reader since they do not differ significantly from what we have done here.

\section{Isospectral hamiltonians arising from the VCS}

In this section we will construct several examples of intertwining operators and their associated hamiltonians using as main ingredient the operator $\tilde A_\gamma$ introduced in (\ref{219}), (\ref{220}), and its adjoint. In a sense we are here reversing the procedure proposed in \cite{bag} where the coherent states were constructed from intertwing operators. Here we have first introduced our VGKCS, and now we will use them to construct pairs of isospectral hamiltonians.

In  \cite{bag} we have shown that if $h$ is a self-adjoint hamiltonian on the Hilbert space $\Hil$, $h=h^\dagger$, whose normalized eigenvectors, $\hat\varphi_n$, satisfy the equation: $h\hat\varphi_n=\epsilon_n\hat\varphi_n$, $n\in\Bbb{N}_0:=\Bbb{N}\cup\{0\}$, and if there exists an operator $x$ such that
$[xx^\dagger,h]=0$,
and  $N_1:=x^\dagger\,x$ is invertible then, calling $H:=N_1^{-1}\left(x^\dagger\,h\,x\right)$, and $\Phi_n=x^\dagger\hat\varphi_n$
the following conditions are satisfied:
$[\alpha]$ $H=H^\dagger$, $[\beta]$ $x^\dagger\left(x\,H-h\,x\right)=0$ and $[\gamma]$
if $\Phi_n\neq 0$  then $H\Phi_n=\epsilon_n\Phi_n$. As we have discussed in the Introduction, this method is an improvement with respect to the previously existing literature since we can explicitly construct a new hamiltonian, $H$, which is isospectral to $h$ and whose eigenvectors are related to those of $h$. Now we will show that, working in the assumptions of Section II.1, it is possible to produce pairs of isospectral hamiltonians acting on $\Hil_{susy}$. For that, and also in view of extension to higher dimensional systems,  it is convenient to modify a little bit the notation used so far, avoiding the use of the suffixes $(b)$ and $(f)$. Let then $h_1$ and $h_2$ be two self-adjoint hamiltonians with EDS, and let $\{\varphi_n^{(j)}, \,n\in\Bbb{N}_0, \,j=1,2\}$ be their related eigenvectors: $h_j\varphi_n^{(j)}=\epsilon_n^{(j)}\,\varphi_n^{(j)}$, $n=0,1,2,\ldots$ and $j=1,2$. We assume that $0<\epsilon_0^{(j)}<\epsilon_1^{(j)}<\epsilon_2^{(j)}<\cdots$, $j=1,2$, and we define
$$
\tilde\epsilon_n^{(j)}=\epsilon_n^{(j)}-\epsilon_0^{(j)}, \quad\mbox{ so that }\quad \tilde\epsilon_0^{(j)}!=1, \quad \tilde\epsilon_n^{(j)}!=(\epsilon_n^{(j)}-\epsilon_0^{(j)})\cdots(\epsilon_1^{(j)}-\epsilon_0^{(j)}), \quad\mbox{ if }\quad n>0.
$$
Further we introduce, as in the previous section,
\be
  \hat\Psi_n^{(1)}=\left(
                                     \begin{array}{c}
                                       \hat\varphi_n^{(1)} \\
                                       0 \\
                                     \end{array}
                                   \right), \quad \hat\Psi_n^{(2)}=\left(
                                     \begin{array}{c}
                                       0 \\
                                       \hat\varphi_n^{(2)} \\
                                     \end{array}
                                   \right),
                                   \label{a1}\en
and
\be
\Psi(\underline J,\gamma):=\frac{1}{\sqrt{\tilde N(\underline J)}}\sum_{n=0}^\infty\,\,\left(\frac{1}{\sqrt{\tilde\epsilon_n^{(1)}!}}J_1^{n/2}\,e^{-i\epsilon_n^{(1)}\gamma}\,\hat\Psi_n^{(1)}+
\frac{1}{\sqrt{\tilde \epsilon_n^{(2)}!}}J_2^{n/2}\,e^{-i\epsilon_n^{(2)}\gamma}\,
\hat\Psi_n^{(2)}\right).
\label{a2}\en
As we have shown  previously these states are VCS, satisfying all the properties which are usually required in the Gazeau-Klauder settings. In particular, what we need now is the fact that they are eigenstates of an annihilation-like operator, see (\ref{219}), which we now simply define as
\be
B_\gamma\hat\Psi_n^{(j)}=
\sqrt{\tilde\epsilon_n^{(j)}}\,e^{i(\epsilon_n^{(j)}-\epsilon_{n-1}^{(j)})\gamma}\hat\Psi_{n-1}^{(j)},
\label{a3}\en
$j=1,2$, $n\in\Bbb{N}_0$, understanding that, since $\tilde\epsilon_0^{(j)}=0$ for $j=1,2$, the action of $B_\gamma$ on  $\hat\Psi_0^{(j)}$ returns the zero vector. The adjoint of $B_\gamma$ is
\be
B_\gamma^\dagger\hat\Psi_n^{(j)}=
\sqrt{\tilde\epsilon_{n+1}^{(j)}}\,e^{-i(\epsilon_{n+1}^{(j)}-\epsilon_{n}^{(j)})\gamma}\hat\Psi_{n+1}^{(j)},
\label{a4}\en
$j=1,2$. As we have already seen, $B_\gamma\Psi(\underline J,\gamma)=J^{1/2}\Psi(\underline J,\gamma)$: for each fixed $\gamma$ the vector $\Psi(\underline J,\gamma)$ is an eigenstate of the operator $B_\gamma$. It should be noticed however that $B_\gamma\Psi(\underline J,\gamma')\neq\,J^{1/2}\Psi(\underline J,\gamma')$. These annihilation-like operators will be used to construct our examples.

\vspace{2mm}

{\bf Remark:--} of course the explicit expression for $B_\gamma$ depends on the vectors $\hat\Psi_n^{(j)}$ and on the sequences $\epsilon_{n}^{(j)}$, $j=1,2$. A simple example can be constructed starting from two harmonic oscillators: let $h_1=\omega_1 a_1^\dagger a_1$ and $h_2=\omega_2 a_2^\dagger a_2$, with $\omega_j>0$, $j=1,2$, and  $[a_i,a_j^\dagger]=\delta_{i,j}\Id$. Hence $\varphi_n^{(j)}=\frac{1}{\sqrt{n!}}\,(a_j^\dagger)^n\varphi_0^{(j)}$, where $a_j\varphi_0^{(j)}=0$, and $\epsilon_{n}^{(j)}=\omega_j\,n$, $j=1,2$ and $n\in\Bbb{N}_0$. Then $\hat\Psi_n^{(1)}$ and $\hat\Psi_n^{(2)}$ can be easily found from (\ref{a1}) and
$$
B_\gamma = \left(
        \begin{array}{cc}
         \sqrt{\omega_1} e^{i\omega_1\gamma}\,a_1 & 0 \\
          0 & \sqrt{\omega_2} e^{i\omega_2\gamma}\,a_2 \\
        \end{array}
      \right).
$$

\vspace{2mm}

{\bf Example 1:--} Let us define the  self-adjoint operator $h_\gamma:=B_\gamma^\dagger\,B_\gamma$ on $\Hil_{susy}$. The set $\F=\{\hat\Psi_n^{(j)},\, n\geq 0,\,j=1,2\}$ is an orthonormal set of eigenvectors of $h_\gamma$: $h_\gamma\,\hat\Psi_n^{(j)}=\tilde\epsilon_{n}^{(j)}\,\hat\Psi_n^{(j)}$, $n\in\Bbb{N}_0$ and $j=1,2$. Hence the operator $h_\gamma$ turns out to be independent of $\gamma$. For this reason, quite often from now on, we will call it simply $h$. Let us now take $x\rightarrow x_\gamma:=B_\gamma^\dagger$. It is clear that $[h,x_\gamma x_\gamma^\dagger]=[B_\gamma^\dagger\,B_\gamma,B_\gamma^\dagger\,B_\gamma]=0$. Furthermore, since $N_1\hat\Psi_n^{(j)}=\tilde\epsilon_{n+1}^{(j)}\hat\Psi_n^{(j)}$, and since $\tilde\epsilon_{n+1}^{(j)}>0$ for all $n\in\Bbb{N}_0$,  $j=1,2$, $N_1^{-1}$ exists and is defined on the orthonormal basis $\F$ of $\Hil_{susy}$ as $N_1^{-1}\hat\Psi_n^{(j)}=(\tilde\epsilon_{n+1}^{(j)})^{-1}\,\hat\Psi_n^{(j)}$. Then all the requirements in \cite{bag} are satisfied, and we find that the operator $H_\gamma=N_1^{-1}\left(x_\gamma^\dagger\, h\,x_\gamma\right)=B_\gamma\,B_\gamma^\dagger$ is isospectral to $h$. Again $H_\gamma$ does not depend on $\gamma$. Its eigenvectors are $\Phi_n^{(j)}=x_\gamma^\dagger\hat\Psi_n^{(j)}=B_\gamma\hat\Psi_n^{(j)}$, which are different from zero if $n\geq1$. This example is nothing but ordinary SUSY quantum mechanics.

\vspace{2mm}

{\bf Example 2:--} Let us again define $h_\gamma\rightarrow h:=B_\gamma^\dagger\,B_\gamma$.  We take $x_\gamma:=\left(B_\gamma^\dagger\right)^2$. Once again we can check that the operator $N_1=B_\gamma^2\,\left(B_\gamma^\dagger\right)^2$ can be inverted, since $N_1\hat\Psi_n^{(j)}=\tilde\epsilon_{n+1}^{(j)}\tilde\epsilon_{n+2}^{(j)}\hat\Psi_n^{(j)}$, noticing that $\tilde\epsilon_{n+1}^{(j)}\tilde\epsilon_{n+2}^{(j)}>0$ for all $n\in\Bbb{N}_0$ and for $j=1,2$. Moreover $h$ commutes with $x_\gamma x_\gamma^\dagger$: $[h,x_\gamma x_\gamma^\dagger]=[B_\gamma^\dagger\,B_\gamma,\left(B_\gamma^\dagger\right)^2\,B_\gamma^2]=0$. Hence we can produce a second hamiltonian, $H_\gamma$, with the same set of eigenvalues as $h$: $H_\gamma=\left(B_\gamma^2\left(B_\gamma^\dagger\right)^2\right)^{-1}\left(B_\gamma^2\,B_\gamma^\dagger\,B_\gamma\,
\left(B_\gamma^\dagger\right)^2\right)$. Here a new feature appears:  in all the examples considered in \cite{bag}, as well as in the example above, the operator $N_1^{-1}$ disappears from the final expression of the companion hamiltonian $H$ of $h$, since is always right-multiplied by $N_1$. In the present case, this is not so. Indeed, we cannot simplify much further the above expression for $H_\gamma$, so that $N_1^{-1}$ appears in the final result, contrarily to what was suggested in \cite{bag}. Nevertheless we can still explicitly check that the set of vectors $\Phi_n^{(j)}=x_\gamma^\dagger\hat\Psi_n^{(j)}=B_\gamma^2\hat\Psi_n^{(j)}$, which is different from zero if $n\geq2$, is a set of eigenvectors of $H_\gamma$, with the same eigenvalues of $h$, $\tilde\epsilon_{n}^{(j)}$. Indeed we find
$$
H_\gamma\left(B_\gamma^2\hat\Psi_n^{(j)}\right)=\tilde\epsilon_{n}^{(j)}\,\left(B_\gamma^2\hat\Psi_n^{(j)}\right),
$$
for all $n\geq2$ and $j=1,2$.

\vspace{2mm}

{\bf Example 3:--} The above examples can be further generalized to other powers of $B_\gamma^\dagger$ in $x_\gamma$. In particular, if we take $h:=B_\gamma^\dagger\,B_\gamma$ and $x_\gamma:=\left(B_\gamma^\dagger\right)^3$, it is easy to check that $h$ commutes with $x_\gamma x_\gamma^\dagger$ and that $N_1=x_\gamma^\dagger x_\gamma$ is invertible. Then our strategy can be applied and we deduce that the vectors $B_\gamma^3\Psi_n^{(j)}$, $n\geq3$ and $j=1,2$, are eigenstates of $H_\gamma=\left(B_\gamma^3\left(B_\gamma^\dagger\right)^3\right)^{-1}\left(B_\gamma^3\,B_\gamma^\dagger\,B_\gamma\,
\left(B_\gamma^\dagger\right)^3\right)$ with the same eigenvalues $\tilde\epsilon_{n}^{(j)}$ as $h$. The extension to $x_\gamma:=\left(B_\gamma^\dagger\right)^l$, $l\geq4$, is straightforward and will not be given here.

\vspace{2mm}

{\bf Example 4:--} The same operators $B_\gamma$ and $B_\gamma^\dagger$ can be used to construct an example of different kind. Let us take $h_\gamma:={B_\gamma^\dagger}^2\,B_\gamma^2$ and $x_\gamma:=B_\gamma^\dagger$. The eigenstates of $h$ are {\em almost} the same as before. More explicitly we have, for all $n\geq 2$, $h_\gamma\Psi_n^{(j)}=\tilde\epsilon_{n}^{(j)}\tilde\epsilon_{n-1}^{(j)}\,\Psi_n^{(j)}$. Hence, $h_\gamma$ does not depend on $\gamma$, and can be simply called $h$. Once again, it is easy to check that $h$ commutes with $x_\gamma x_\gamma^\dagger$ and that $N_1=x_\gamma^\dagger x_\gamma$ is invertible. In this case $H_\gamma$ turns out to be the operator $H_\gamma=B_\gamma^\dagger B_\gamma^2 B_\gamma^\dagger$, and the eigenstates can be obtained as $B_\gamma\Psi_n^{(j)}$. The eigenvalues, as always, are the same as those of $h$.

\vspace{2mm}

{\bf Remark:--} We notice that all has been discussed in this section can be extended to a set of $N$ hamiltonians $h_j$, $j=1,2,\ldots,N$ with EDS. In this case, the relevant Hilbert space would be $\Hil_{susy}=\Bbb{C}^N\otimes\Hil$. Furthermore, changing our definitions a little bit, we claim that more or less the same results could be obtained starting from a set of hamiltonians even if they do not have EDS.

\vspace{2mm}

We end this section showing the link between the {\em original hamiltonians} $h_1,h_2$ and some of the operators $h_\gamma$ introduced in the examples above. Using the matrix operator $H_\tau$ introduced in the previous section we find that $H_\tau\hat\Psi_n^{(j)}=\tilde\epsilon_n^{(j)}\hat\Psi_n^{(j)}$, and we have already seen that $B_\gamma^\dagger\,B_\gamma\hat\Psi_n^{(j)}=\tilde\epsilon_n^{(j)}\hat\Psi_n^{(j)}$. This means that $H_\tau-B_\gamma^\dagger\,B_\gamma$ is zero on every vector of a basis of $\Hil_{susy}$, and so $H_\tau=B_\gamma^\dagger\,B_\gamma$. This is an explicit example of a general result in functional analysis which states (but for some mathematical details) that every positive operator $T$ can be written as $T=W^\dagger\,W$, for some operator $W$. We want to stress that here $B_\gamma$ cannot just be taken as the square root of $H_\tau$, since otherwise we would lose one of the main feature of our framework, namely the fact that $\Psi(\underline J,\gamma)$ is an eigenstate of $B_\gamma$.

\section{Non-isospectral hamiltonians: is this an extension?}

We devote this short section to a possible generalization of what has been done in \cite{bag}. We call it {\em possible} because, under suitable conditions, what we are going to discuss here turns out to be equivalent to the results in \cite{bag}. The main idea is to produce, starting from a given hamiltonian $h$, a second operator, $H$ whose spectrum $\sigma(H)$ is different from $\sigma(h)$ but whose eigenstates are related to those of $h$ by means of the usual intertwining operator. As a matter of fact, this is not an easy task using the standard results on intertwining operators, while is just a very simple exercise adopting the strategy in \cite{bag}. Indeed, let  $h$ be a self-adjoint hamiltonian on the Hilbert space $\Hil$, $h=h^\dagger$, whose normalized eigenvectors, $\hat\varphi_n$, satisfy the  equation: $h\,\hat\varphi_n=e_n\hat\varphi_n$, $n\in\Bbb{N}_0:=\Bbb{N}\cup\{0\}$. Suppose that there exists an operator $x$ such that
$[xx^\dagger,h]=0$
and  $N_1:=x^\dagger\,x$ is invertible. So we are back to the hypotheses of the previous section. Let further $f(x)$ be a real $C^\infty$-function which admits a power series expansion. Since $N_1^{-1}$ exists by assumption, we can introduce
\be
H:=N_1^{-1}\left(x^\dagger\,f(h)\,x\right),\qquad \Phi_n=x^\dagger\hat\varphi_n.
\label{32}\en
Here $f(h)$ can be defined, for instance, via functional calculus or, at least on a suitable domain of vectors, considering its power series expansion. Then the following conditions are satisfied:
$[\alpha]$ $H=H^\dagger$;
$[\beta]$ $x^\dagger\left(x\,H-f(h)\,x\right)=0$;
$[\gamma]$ if $\Phi_n\neq 0$ then  $H\Phi_n=E_n\Phi_n$, with $E_n=f(e_n)$.

The proof of these statements does not differ significantly from that given in \cite{bag}, and will not be given here. We want to remark that the hamiltonians $H$ and $h$ are no longer isospectral. On the contrary, choosing $f$ in a clever way, it will turn out that $h$ and $H$ have EDS so that they could be used in the construction of the VGKCS (\ref{214}).

\vspace{2mm}

{\bf Remark:--} it may be worth remarking that, as in \cite{bag}, we could iterate our procedure constructing an entire family of hamiltonians with their related eigensystems.

\vspace{2mm}

Now we came to the title of this section: is the one proposed here really an extension of the results in \cite{bag}? Or is it just another way to say the same thing? In other words, given $\varphi\in\Hil$, does the equality
\be
f\left(N_1^{-1}(x^\dagger\, h\,x)\right)\varphi=N_1^{-1}(x^\dagger\, f(h)\,x)\varphi
\label{33}\en
holds true? If this is the case, then we are taken back to \cite{bag}. On the contrary, if this is not so, then the strategy is really new.  As a matter of fact, we have not a final answer to this problem but many strong indications.

We begin discussing a sufficient condition for (\ref{33}) to be satisfied. Let us now assume that for $l=0,1,2,\ldots$ and for $\varphi\in\Hil$, \be\left(x\,N_1^{-1}x^\dagger\right)h^l\,x\,\varphi=h^l\,x\,\varphi.\label{33b}\en Then, if $f(x)$ admits a power series expansion, (\ref{33}) is satisfied. This can be proven by a simple direct computation. Moreover, if $[x\,N_1^{-1}x^\dagger,h]=0$, then equation (\ref{33b}) is surely satisfied. So, whenever $h$ commutes with $x\,N_1^{-1}x^\dagger$, (\ref{33}) is satisfied. The next step is to check if, for some reason, $[x\,N_1^{-1}x^\dagger,h]=0$ is always satisfied. Indeed, from our definitions, we can conclude that, at least, $x^\dagger\,[x\,N_1^{-1}x^\dagger,h]x=0$. Now, if the range of the operator $x$, $Ran(x)$, is all of $\Hil$, as it happens if $x$ is invertible, then $[x\,N_1^{-1}x^\dagger,h]=0$. However this conclusion does not hold if $Ran(x)$ is a proper subset of $\Hil$, so that nothing can be said about equation (\ref{33}). Summarizing we have:

\begin{prop}
Let $h$ be a self-adjoint operator and $x$ be such that $[h,xx^\dagger]=0$, $N_1:=x^\dagger\,x$ is invertible and $Ran(x)=\Hil$. Then, if the function $f(t)$ admits a power series expansion, equality (\ref{33}) holds.
\end{prop}

However the  examples we are going to discuss, and which are suggested by  bosons, quons and ordinary SUSY quantum mechanics, show that, even in some case in which  $Ran(x)\subset\Hil$, equation (\ref{33}) can still be recovered. As a matter of fact, during our analysis we have not found any example showing that (\ref{33}) is {\bf not} satisfied, so that we cannot give any conclusion at the present stage.

\vspace{3mm}

{\bf Bosons:--}
Let $h=a^\dagger a=:N$, $x=\left(a^\dagger\right)^2$ and $[a,a^\dagger]=\Id$, as in Example 2, \cite{bag}. We know that $[h,xx^\dagger]=0$ and that $N_1=x^\dagger x=N^2+3N+2\Id$ is invertible. In \cite{bag} we have deduced that $N_1^{-1}(x^\dagger\,h\, x)=N+2\Id$ Now, if we fix $f(x)=x^2$ and we compute $H=N_1^{-1}(x^\dagger h^2 x)$, we get $H=\1+aa^\dagger+N_1=(N+2\Id)^2$. If we rather take $f(x)=e^x$ we find $H=e^{N+2\Id}$. More in general, if $f(x)$ admits a power series expansion, we find that $H=f(N+2\Id)$. Hence (\ref{33}) is satisfied, even if $Ran(x)\subset\Hil$. This inclusion follows because, calling $\varphi_0$ the vacuum of $a$, $a\varphi_0=0$, then both $\varphi_0$ and $a^\dagger\varphi_0$ are orthogonal to $Ran(x)$.

\vspace{3mm}

{\bf Quons:--}
Once again we take $h=a^\dagger a=:N$ and $x=\left(a^\dagger\right)^2$, but we consider the following q-mutation relation: $aa^\dagger-qa^\dagger a=\Id$, as in \cite{moh}. We know, \cite{bag}, that $[h,xx^\dagger]=0$ and that $N_1=x^\dagger x=q^3N^2+q(1+2q)N+(1+q)\Id$ is invertible, at least if $0<q\leq1$. In \cite{bag} we have obtained that $N_1^{-1}(x^\dagger\,h\, x)=(1+q)\Id+q^2\,N$  Once again, if $f(x)$ admits a power series-expansion, we  get $H=N_1^{-1}(x^\dagger f(h) x)=f\left(q^2N+(1+q)\Id\right)$, even if $Ran(x)$ is again a proper subset of $\Hil$.

\vspace{3mm}

{\bf Susy quantum mechanics:--}
We consider now the usual hamiltonian $h=a^\dagger a=:N$, but we take $x=a^\dagger$ and $[a,a^\dagger]=\frac{2\hbar}{\sqrt{2m}}W'(x)$, where $W(x)$ is the so-called superpotential, $\hbar$ is the Plank constant (divided by $2\pi$) and $m$ is the mass of a certain quantum particle. If we assume that $W'(x)>0$ for all $x$ it is possible to check that our assumptions are all satisfied. If we further take a generic function $f(x)$, which admits a power series expansion, then we get $H=f\left(a^\dagger a+\frac{2\hbar}{\sqrt{2m}}W'(x)\right)$. Even in this example equation (\ref{33}) is recovered regardless of the fact that $Ran(x)$ may be a proper subset of $\Hil$.

Different choices of the operator $x$ trivializes the situation. If we take, for instance, $x=a$ or $x=\left(a^\dagger\right)^2$, we deduce that, in order to satisfy our assumptions,  $W(x)$ must be linear in $x$, and this produces the standard harmonic oscillator, which is not very interesting.

\vspace{3mm}

We are therefore left with an open problem: since many choices of $x$ can be done such that $Ran(x)\subset\Hil$, we wonder whether in some of these cases equation (\ref{33}) is not verified. We hope to be able to give a final answer in a close future.

\section{Conclusions}

In this paper we have continued our analysis on SUSY quantum mechanics, intertwining operators and coherent states began in \cite{bag}. As the reader can see, different definitions of coherent states are given here, in \cite{bag} and in other papers on the same subject, see \cite{fhro,aleixo} for instance. In our opinion there is no reasonable way to decide which is the {\em best} definition, at least until no concrete physical application is considered. In other words: a given definition of coherent state may be useful for a particular application but not for a different one. No other general rule does exist.

Also, in Section IV we have given some preliminary results on non-isospectral hamiltonians. This topic surely deserves a deeper analysis, also in connection with interesting alternatives which already exist in the literature, \cite{spi,fern}, where possible generalizations on the intertwining operators are considered. The work in \cite{spi,fern} seems also to be connected with the examples considered here involving quons, and we plan to consider in more details also this aspect.

\section*{Acknowledgements}

This work was supported by M.U.R.S.T..

\end{document}